\documentclass[twocolumn]{aastex701}

\usepackage{multirow}
\usepackage{booktabs}
\usepackage{natbib}
\usepackage{graphicx}
\usepackage{amsmath}
\usepackage{textgreek}
\usepackage{graphicx}
\usepackage{soul}
\usepackage{subcaption}
\usepackage{array}
\usepackage{tabularx}
\usepackage{float}
\usepackage{makecell}
\usepackage{color}
\usepackage{hyperref}
\begin{document}

\title{Barred Galaxies in MaNGA: Stellar Mass Dependence and Black Hole Growth}

\author[orcid=0009-0005-2923-9933,gname='Chandan',sname='Watts']{Chandan Watts}
\affiliation{Indian Institute of Astrophysics,II Block, Koramangala, Bengaluru 560034, INDIA.}
\affiliation{Pondicherry University, R.V. Nagar, Kalapet, 605014, Puducherry, India}
\email[show]{chandan@iiap.res.in, chandanwatts510@gmail.com}  

\author[orcid=0000-0002-3927-5402]{Sudhanshu Barway}
\affiliation{Indian Institute of Astrophysics,II Block, Koramangala, Bengaluru 560034, INDIA.}
\affiliation{Pondicherry University, R.V. Nagar, Kalapet, 605014, Puducherry, India}
\email{sudhanshu.barway@iiap.res.in}

\begin{abstract}
This study investigates the role of internal structures, such as stellar bars and active galactic nuclei (AGN), in shaping galaxy evolution using 7,408 galaxies (2,641 barred) from the SDSS-IV MaNGA survey. Our analysis spans a wide stellar mass range, from dwarf to massive galaxies, examining how internal and environmental processes regulate galaxy evolution. Galaxies are categorized by morphology (ellipticals (E), lenticulars (S0s), early-type spirals (ETS), and late-type spirals (LTS)), environment, and the presence of a bar. We find that barred galaxies have higher median specific star formation rate (sSFR) than unbarred galaxies across most stellar mass intervals. This difference varies with host morphology, with LTS and S0 galaxies showing different trends across stellar mass intervals. The environmental trends of barred and unbarred galaxies are broadly similar within the same mass intervals. In AGN hosts, barred Seyferts tend to lie near the star-forming–green valley boundary, while barred LINERs preferentially occupy the green valley.  We further examine the relation between stellar bars and black hole mass and find that barred galaxies host systematically lower black hole masses than unbarred galaxies, with environmental effects appearing to play a secondary role in shaping these differences. A multivariate regression analysis also shows that the positive bar--sSFR association remains statistically significant for star-forming galaxies after accounting for stellar mass, local density, AGN activity, and morphology. Our results highlight the interconnected roles of stellar bars, stellar mass, morphology, and environment in shaping galaxy star formation, providing new insights into the role of stellar bars in galaxy evolution.
\end{abstract}


\section{Introduction} 
Bars are among the most influential internal structures in galaxies, efficiently channeling gas toward the center to trigger nuclear star formation or fuel AGN activity \citep{Sheth_2005, 2011MNRAS.416.2182E, 2011MNRAS.411.2026M}. However, sustained inflows may also deplete disk gas, suppressing outer-disk star formation \citep{kruk2018,  Fraser_2020, Geron_2021, Geron_2023}. In addition to internal processes, barred galaxies are also affected by external environmental factors. Previous studies have shown that galaxy properties vary significantly with local density and within cluster environments \citep{2009A&A...497..713B,2014MNRAS.445.1339L,2016ApJ...826..227L,2018ApJ...857....6L,2025A&A...702A...7L,2026A&A...705A.115C}. Environmental mechanisms such as tidal interactions and mergers can significantly modify galaxy structure and star-formation activity. 

Active galactic nuclei (AGN) feedback serves as another key regulator, injecting energy into the interstellar medium through radiation, winds, or jets, and either suppressing or stimulating star formation depending on accretion state \citep{Zubovas_2013, Harrison_2017, Zhuang_2021}. Disentangling the interactions between AGN activity, host morphology, and environment is crucial for understanding galaxy–black hole co-evolution. Observational studies have shown that supermassive black hole mass correlates tightly with the properties of classical bulges and elliptical galaxies, indicating a co-evolution between galaxies and their central black holes \citep{2013ARA&A..51..511K, 2013ApJ...764..184M}. However, these correlations vary with galaxy morphology and are significantly weaker in galaxies hosting pseudobulges, suggesting that black hole growth pathways may differ across morphological types.

\begin{figure}[htbp]
    \centering
    \includegraphics[width=\columnwidth]{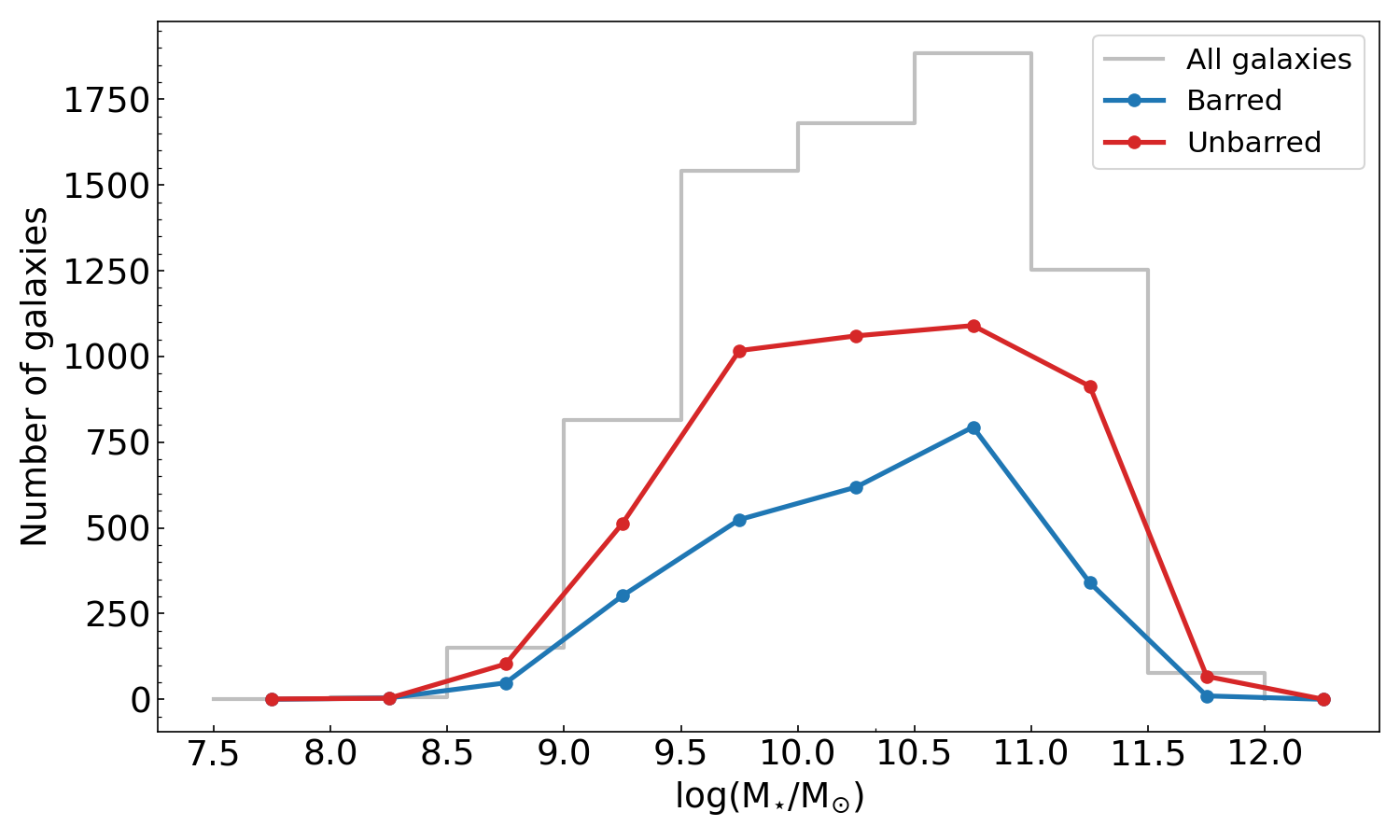}
    \caption{Number of barred (blue) and unbarred (red) galaxies in the stellar mass bins of 0.5 dex, with the full galaxy sample shown in grey as a step plot.}
    \label{fig:sSFR_scatter_full_sample}
\end{figure}

This paper is part of a series following \citet{2026ApJ..1004..199W}, in which we examined the role of external mechanisms, such as local environmental density and galaxy morphology, in shaping star formation within the Mapping Nearby Galaxies at Apache Point Observatory (MaNGA) survey. That study showed that for the galaxies in the MaNGA survey, stellar mass is the primary parameter governing galaxy properties, with dwarf, intermediate-, and high-mass systems exhibiting distinct evolutionary trends. In the present work, we shift our focus to internal processes, specifically the presence of AGN and stellar bars, and examine how these components are associated with galaxy properties and how they shape galaxy evolution. 

In this study, we analyze a sample of 7,408 galaxies (2,641 barred and 4,767 unbarred) from the MaNGA survey, spanning a wide range of stellar masses from dwarf to massive galaxies. Using measurements of morphology, bar presence, specific star formation rate (sSFR), AGN activity, and local environmental density, we investigate the statistical trends among these properties. In particular, we examine how internal structures, such as bars and AGN, are associated with galaxy properties across the MaNGA sample.

The paper is organized as follows: Section~\ref{sec:sample} describes the data and sample construction; Section~\ref{sec:results} presents results across mass regimes and AGN hosts; and Section~\ref{sec:conclusion} summarizes our key findings and implications for the coupled evolution of morphology, star formation, and environment.

\begin{table*}
\centering
\caption{Median specific star formation rates (sSFR) of barred and unbarred galaxies in different mass ranges. The median values are accompanied by their bootstrap uncertainties. $\Delta$Median is defined as the difference between the median log(sSFR) of barred and unbarred galaxies (Barred$-$Unbarred). Positive values indicate higher median sSFR in barred galaxies. Elliptical galaxies are excluded from this analysis.}
\label{tab:ssfr_nomorph}
\small
\begin{tabular}{lccccc}
\toprule
\textbf{Mass Range} & \textbf{Bar} & \textbf{No. of galaxies} & \textbf{Median log(sSFR)} & \textbf{$\Delta$Median (dex)} \\

\midrule
\textbf{$<$10 (Overall)} & Barred & 878 & -10.023$\pm$0.018 & 0.123$\pm$0.022\\
      & Unbarred & 1481 & -10.146$\pm$0.015 & \\

\textbf{$\leq$9.5} & Barred & 354 & -9.948$\pm$0.014 & 0.066$\pm$0.023 \\
            & Unbarred & 567 & -10.014$\pm$0.018 &  \\
\textbf{9.5--10} & Barred & 524 & -10.080$\pm$0.018 & 0.161$\pm$0.024 \\
         & Unbarred & 914 & -10.240$\pm$0.016 & \\
\addlinespace
\midrule
\addlinespace
\textbf{$\geq$10 (Overall)} & Barred & 1763 & -10.793$\pm$0.021 & 0.149$\pm$0.032 \\
          & Unbarred & 2311 & -10.942$\pm$0.024 &  \\

\textbf{10--10.5} & Barred & 619 & -10.431$\pm$0.026 & 0.125$\pm$0.041 \\
          & Unbarred & 919 & -10.556$\pm$0.035 &  \\

\textbf{10.5--11} & Barred & 794 & -10.831$\pm$0.048 & 0.146$\pm$0.064 \\
          & Unbarred & 862 & -10.978$\pm$0.043 &  \\

\textbf{$\geq$11} & Barred & 350 & -10.989$\pm$0.021 & 0.439$\pm$0.074 \\
          & Unbarred & 530 & -11.428$\pm$0.071 &  \\
\bottomrule
\end{tabular}
\end{table*}

\section{Data and Sample Description} \label{sec:sample}

The master sample is constructed by combining four major catalogs: the SDSS--MaNGA survey Data Release 17 (DR17; \citealt{2015ApJ...798....7B, 2017AJ....154...86W}), the MaNGA Visual Morphology Catalog \citep{2022MNRAS.512.2222V}, the GALEX--SDSS--WISE Legacy Catalog (GSWLC; \citealt{2016ApJS..227....2S, 2018ApJ...859...11S}), and the Environmental Density Catalog \citep{2006MNRAS.373..469B}. A detailed description of these catalogs and the cross-matching procedure is provided in \citet{2026ApJ..1004..199W}. We cross-match the MaNGA catalog with the morphology catalog to obtain visual classifications, with GSWLC for stellar mass and star formation rates, and with the environmental density catalog for local density measurements.

Galaxies are grouped into four morphological classes: Ellipticals (E), Lenticulars (S0s: S0, S0/a), Early-Type Spirals (ETS: Sa, Sab, Sb, Sbc), and Late-Type Spirals (LTS: Sc, Scd, Sd, Sdm, Sm, Im, S). The final master sample consists of 7,408 galaxies with complete morphological, stellar mass, star formation rate, and environmental information.

AGN classifications are taken from \citet{2023A&A...674A..85A}, who used SDSS--MaNGA DR17 data and BPT diagnostics with multiple aperture definitions. Their catalog includes classifications into Seyfert, LINER, Composite, Star-Forming, and Ambiguous categories, with an EW(H$\alpha$) $> 3$~\AA\ cut applied for the final AGN selection. From their catalog of 10,242 galaxies, 467 are classified as AGNs. Cross-matching with our master sample yields a final AGN sample of 296 galaxies.

Bar classifications are based on the \texttt{BARS} column in the MaNGA Visual Morphology Catalog, which provides the probability of a galaxy hosting a bar. Galaxies with bar probabilities of $\geq 0.5$ are classified as barred, corresponding to galaxies with visually identified bars of varying prominence (0.5: inner bar, 0.75: prominent bar, 1.0: strong bar). Using this criterion, 2,641 galaxies in the master sample are classified as barred.

We classify galaxies into three environmental regimes following \citet{2006MNRAS.373..469B}: low-density ($\log \Sigma \, (\mathrm{Mpc^{-2}}) \leq -0.5$), intermediate-density ($-0.5 < \log \Sigma \, (\mathrm{Mpc^{-2}}) < 0.5$), and high-density ($\log \Sigma \, (\mathrm{Mpc^{-2}}) \geq 0.5$). Using the classification scheme of \citet{2014SerAJ.189....1S}, galaxies are further separated based on $\log(\mathrm{sSFR})$ into star-forming ($\log(\mathrm{sSFR}) \geq -10.8$), green valley ($-10.8 < \log(\mathrm{sSFR}) < -11.8$), and quenched ($\log(\mathrm{sSFR}) \leq -11.8$) galaxies.

The stellar mass distribution of barred and unbarred galaxies in the MaNGA sample is shown in Figure~\ref{fig:sSFR_scatter_full_sample}. The sample comprises 2,641 barred and 4,767 unbarred galaxies, spanning a wide stellar mass range. The total numbers of barred and unbarred galaxies in the different stellar mass regimes are listed in Table~\ref{tab:ssfr_nomorph}. This mass-based division enables a systematic comparison of how morphology, star formation, and local environmental density vary across different stellar mass regimes.

\section{Results and Discussion}\label{sec:results}
In this study, we adopt a statistical approach to investigate how galaxies evolve across these mass regimes by analyzing their morphology, local environmental density, and the potential influence of bars and AGN. In addition, we compare our findings with those of previous studies to place them in a broader evolutionary context and to assess how stellar mass influences the structural and star-forming properties of barred galaxies. The following sections present these results in detail.

\subsection{Different mass-regimes}

Previous work by \citet{2026ApJ..1004..199W} suggested that galaxies in different stellar mass regimes, including dwarf galaxies ($\log(M_\star/M_\odot)\leq9.5$) and intermediate-mass galaxies  ($9.5 < \log(M_\star/M_\odot)< 10$), exhibit distinct evolutionary trends. Motivated by these results, we compare the star-formation properties of barred and unbarred galaxies across 0.5 dex in stellar mass bins. We adopt the mass bins $\log(M_\star/M_\odot)\leq9.5$, $9.5<\log(M_\star/M_\odot)<10$, $10\leq\log(M_\star/M_\odot)<10.5$, $10.5\leq\log(M_\star/M_\odot)<11$, and $\log(M_\star/M_\odot)\geq11$. The lowest and highest mass bins are chosen to ensure a sufficient number of galaxies for a meaningful statistical comparison.

\subsubsection{Star-formation} \label{sec:sf}

Table~\ref{tab:ssfr_nomorph} summarizes the median sSFR of barred and unbarred galaxies together with the difference between their median values ($\Delta$Median). To minimize the influence of fundamentally different evolutionary pathways, only disk galaxies (S0s, ETS, and LTS) are included in this analysis.

For galaxies with $\log(M_\star/M_\odot)<10$, barred galaxies exhibit a median sSFR that is higher by 0.123 dex than unbarred galaxies. When this mass range is further divided, the enhancement is found to be weaker among dwarf galaxies ($\log(M_\star/M_\odot)\leq9.5$) than in the intermediate-mass interval ($9.5<\log(M_\star/M_\odot)<10$),
indicating that the difference between barred and unbarred galaxies is more pronounced in the latter.

A similar analysis is carried out for galaxies with $\log(M_\star/M_\odot)\geq10$. Overall, barred galaxies show a median sSFR enhancement of 0.149 dex relative to unbarred galaxies. Among the individual mass bins, the largest difference is observed for galaxies with $\log(M_\star/M_\odot)\geq11$, where barred galaxies exhibit a median sSFR higher by 0.439 dex. In comparison, the $10\leq\log(M_\star/M_\odot)<10.5$ and $10.5\leq\log(M_\star/M_\odot)<11$ intervals show comparable median differences of 0.125 dex and 0.146 dex, respectively.

Previous observational and theoretical studies report both enhanced and suppressed star formation in barred galaxies. \citet{2011MNRAS.416.2182E} find that barred galaxies with $\log(M_\star/M_\odot)>10$ have higher central SFRs than their unbarred counterparts, with the enhancement being less evident at lower stellar masses. Similarly, hydrodynamical simulations by \citet{2016MNRAS.463.1074C} show that bars can drive substantial gas inflows and enhance star formation in massive barred galaxies, although subsequent gas consumption can lead to a decline in star formation. Using SDSS-MaNGA data, \citet{2020MNRAS.499.1406L} also find that centrally enhanced star formation is predominantly associated with barred galaxies above $\log(M_\star/M_\odot)>10$. These results are broadly consistent with our finding that the difference in global sSFR between barred and unbarred disk galaxies becomes more pronounced at higher stellar masses.

However, other studies suggest that bars can also be associated with reduced star formation activity. \citet{2020MNRAS.495.4158F} find that, at fixed stellar mass and morphology, barred galaxies have older stellar populations and lower H\textsc{I} gas fractions at high stellar masses, linking bars with earlier cessation of star formation. Similarly, \citet{2020ApJ...893...19W} find both enhanced and suppressed central star formation among late-type barred galaxies, indicating that the presence of a bar alone does not necessarily determine the star formation state of its host. The different results may reflect differences in the spatial scale of the SFR measurements, stellar mass, morphology, and the evolutionary stage of the bar. In contrast to studies focused primarily on central star formation, our analysis considers the global sSFR and finds a positive difference between barred and unbarred galaxies, with the largest offset occurring at the highest stellar masses.

\subsubsection{Morphology}
\begin{figure}
    \centering
    \includegraphics[width=\columnwidth]{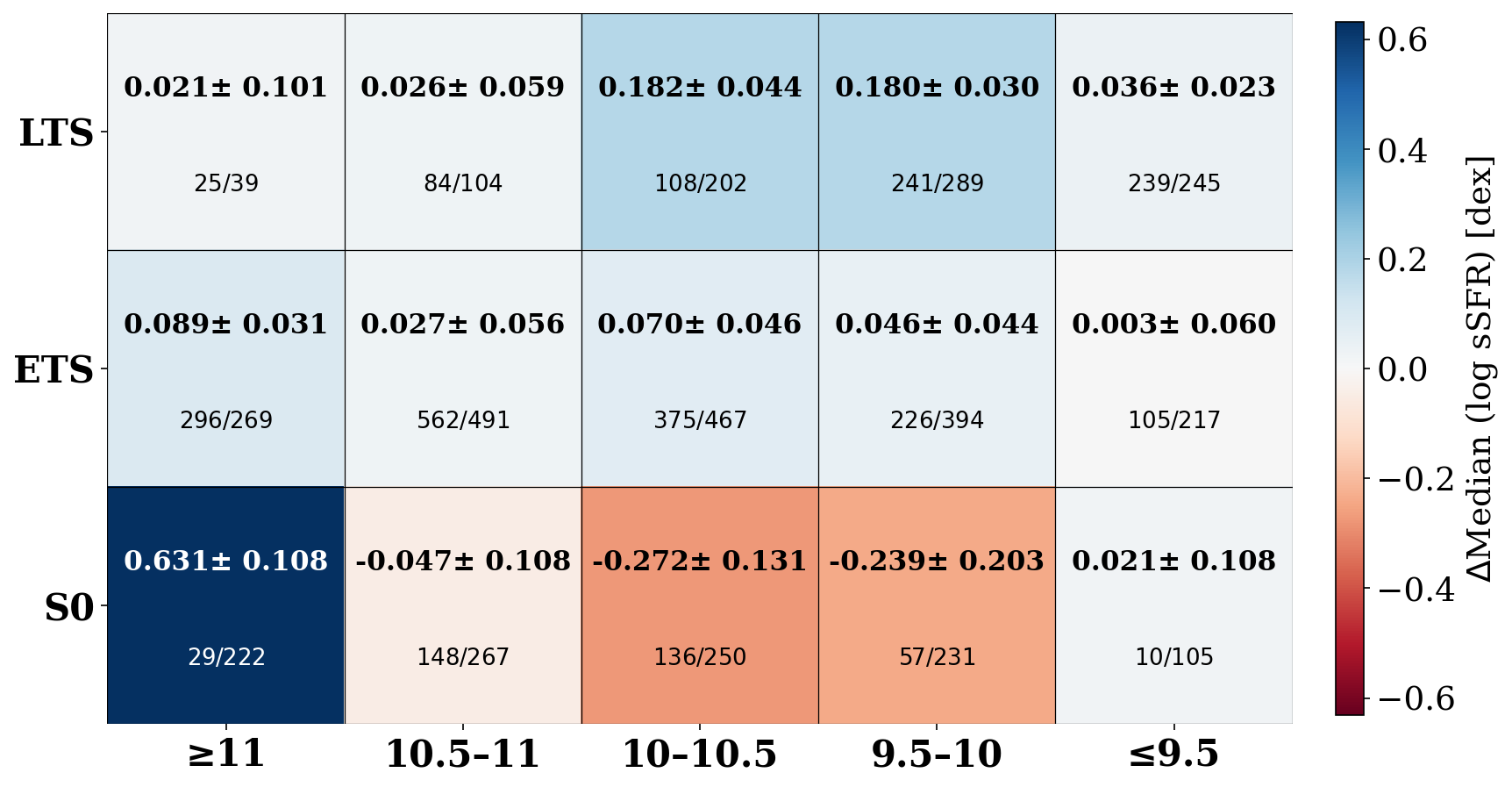}
    \caption{Difference in the median log(sSFR) between barred and unbarred galaxies as a function of stellar mass and morphology. The color of each cell represents $\Delta$Median($\log$ sSFR), where positive (blue) and negative (red) values indicate higher and lower median sSFR in barred galaxies relative to unbarred galaxies, respectively. The value in each cell is given as $\Delta$Median($\log$ sSFR) $\pm$ the bootstrap uncertainty, while the numbers below denote the number of barred and unbarred galaxies (Barred/Unbarred) in that bin.}
    \label{fig:morphology}
\end{figure}

To examine whether the differences in median sSFR depend on galaxy morphology, we compare the barred and unbarred populations within each stellar mass interval for the three morphological classes (LTS, ETS, and S0s). Figure~\ref{fig:morphology} presents the difference in the median sSFR ($\Delta$Median) between barred and unbarred galaxies. The number of galaxies in several morphology--mass bins is relatively small; therefore, these results should be interpreted with caution. The median values with bootstrap uncertainties are adopted to reduce the influence of outliers.

The intermediate-mass interval, $9.5 < \log(M_\star/M_\odot) < 10$, which exhibits the largest overall difference in sSFR between barred and unbarred galaxies (Section~\ref{sec:sf}), is primarily associated with LTS galaxies, which show a positive median difference of 0.18$\pm$0.03 dex. In contrast, S0 galaxies in the same mass range exhibit a lower median sSFR for barred galaxies (-0.24$\pm$0.20 dex), while ETS galaxies show only a small difference (0.05$\pm$0.04 dex).

A similar behavior is found for galaxies with $10 \leq \log(M_\star/M_\odot) < 10.5$, where LTS galaxies continue to exhibit a positive median difference (0.18$\pm$0.04 dex), whereas S0 galaxies show a negative difference (-0.27$\pm$0.13 dex). The ETS population again displays only a modest offset (0.07$\pm$0.04 dex).

For the $10.5 \leq \log(M_\star/M_\odot) < 11$ interval, the median differences are small for all morphological classes, suggesting no clear distinction between barred and unbarred galaxies within this mass range.

In the highest stellar mass bin ($\log(M_\star/M_\odot) \geq 11$), the trend differs from the its lower mass intervals. While LTS and ETS galaxies exhibit only small positive differences (0.02 and 0.09 dex, respectively), S0 galaxies show the largest positive median difference (0.63$\pm$0.10 dex). However, this result is based on a limited number of galaxies. 

The morphology dependence of the bar--star formation relation has also been reported in previous studies. Early observational studies found that the enhancement of star formation associated with bars is more evident in early-type spirals than in late-type spirals \citep{1996A&A...313...13H}. Our morphology-dependent results show a different behavior for the three disk classes. The positive $\Delta$Median values for LTS galaxies in the $9.5<\log(M_\star/M_\odot)<10$ and $10\leq\log(M_\star/M_\odot)<10.5$ intervals contrast with the stronger bar-related star formation reported for early-type spirals by \citet{1996A&A...313...13H}. For S0 galaxies, however, we find negative $\Delta$Median values in these two mass intervals. This is broadly consistent with results from the EAGLE simulations, where barred S0s have lower star formation rates and gas fractions than unbarred S0s, while barred spirals show slightly higher star formation rates than their unbarred counterparts \citep{2022MNRAS.510.5164C}. 

The positive $\Delta$Median observed for S0 galaxies at $\log(M_\star/M_\odot)\geq11$ differs from the trend seen at lower masses and is broadly consistent with the evidence for recent star formation in a subset of barred S0 galaxies reported by \citet{2020MNRAS.495.4548B}. However, our result is based on only 29 barred S0 galaxies and should therefore be interpreted with caution. These comparisons suggest that the relation between bars and star formation is sensitive to both morphology and stellar mass. In particular, the contrasting behavior of spiral and S0 galaxies in our sample indicates that morphology is an important factor to consider when interpreting the difference in sSFR between barred and unbarred galaxies.

\subsubsection{local environemntal density}

We further examine the environmental dependence of barred and unbarred galaxies by comparing the fractions of different morphological and star-formation classes within the barred and unbarred populations as a function of local projected galaxy density ($\log\Sigma$) across different stellar mass intervals. Figure~\ref{fig:bar_morph} presents the environmental dependence of the LTS, ETS, and S0 fractions separately for barred and unbarred galaxies.

For LTS galaxies, the fractions of both barred and unbarred populations generally decrease towards higher local densities across the stellar mass intervals. The separation between the barred and unbarred populations is most evident in the lowest stellar mass bin, while the difference becomes smaller towards higher stellar masses.

The ETS population shows a more varied dependence on local density. In the lowest stellar mass interval ($\log(M_\star/M_\odot)\leq9.5$), the barred ETS fraction increases with density, whereas the unbarred fraction remains relatively stable. In the intermediate-mass intervals, the barred and unbarred fractions show broadly similar variations with density. At higher stellar masses, the barred ETS fraction remains relatively high across the density range, while the unbarred fraction decreases towards higher densities.

For S0 galaxies, the barred and unbarred fractions show broadly similar environmental trends across most stellar mass intervals, with generally modest differences between the two populations. The highest-mass interval also shows relatively small variations with density, although the number of galaxies becomes limited. The overall variation of morphological fractions with local density is consistent with the established morphology--density relation \citep{2009MNRAS.393.1324B, 2023MNRAS.518.5260P}.

We next examine whether the star-formation state of barred and unbarred galaxies shows a similar environmental dependence. Figure~\ref{fig:bar_sf} presents the fractions of star-forming, green valley, and quenched galaxies within the barred and unbarred populations as a function of local projected galaxy density.

Across the stellar mass intervals, the fraction of star-forming galaxies generally decreases with increasing local density for both barred and unbarred populations, consistent with the established star formation--density relation \citep{2010ApJ...721..193P}. The barred and unbarred populations broadly follow similar trends, although the highest-mass interval shows a modest increase in the barred fraction at the highest densities.

The green valley population exhibits a weaker dependence on local density than the star-forming population. The barred and unbarred fractions are broadly comparable across most stellar mass intervals, although the highest stellar mass bin shows a relatively higher barred fraction over much of the density range.

The fraction of quenched galaxies generally increases towards higher local densities for both barred and unbarred populations. The two populations follow similar trends at lower and intermediate masses, while the unbarred fraction becomes higher than the barred fraction in the two highest stellar mass intervals ($\log(M_\star/M_\odot)\geq10.5$).

The apparent differences between barred and unbarred populations in individual density and mass intervals should be interpreted cautiously, as several bins contain relatively few galaxies and the corresponding uncertainties are large.

These results show that stellar mass, morphology, star formation, and local environment are interconnected. The decline in the fraction of star-forming galaxies and the increase in the quenched fraction towards higher densities are accompanied by changes in the relative fractions of LTS, ETS, and S0 galaxies. Thus, the mass- and morphology-dependent differences in sSFR observed above occur alongside systematic variations in local environment. However, barred and unbarred galaxies generally show similar environmental trends within the same stellar mass intervals, suggesting that the observed differences in sSFR between the two populations cannot be attributed solely to variations in local density. The relative contributions of these properties are examined simultaneously using a multivariate regression analysis in Section~\ref{sec:multi}.

\begin{figure*}[t]
    \centering
    \begin{subfigure}{0.46\textwidth}
        \includegraphics[width=\textwidth]{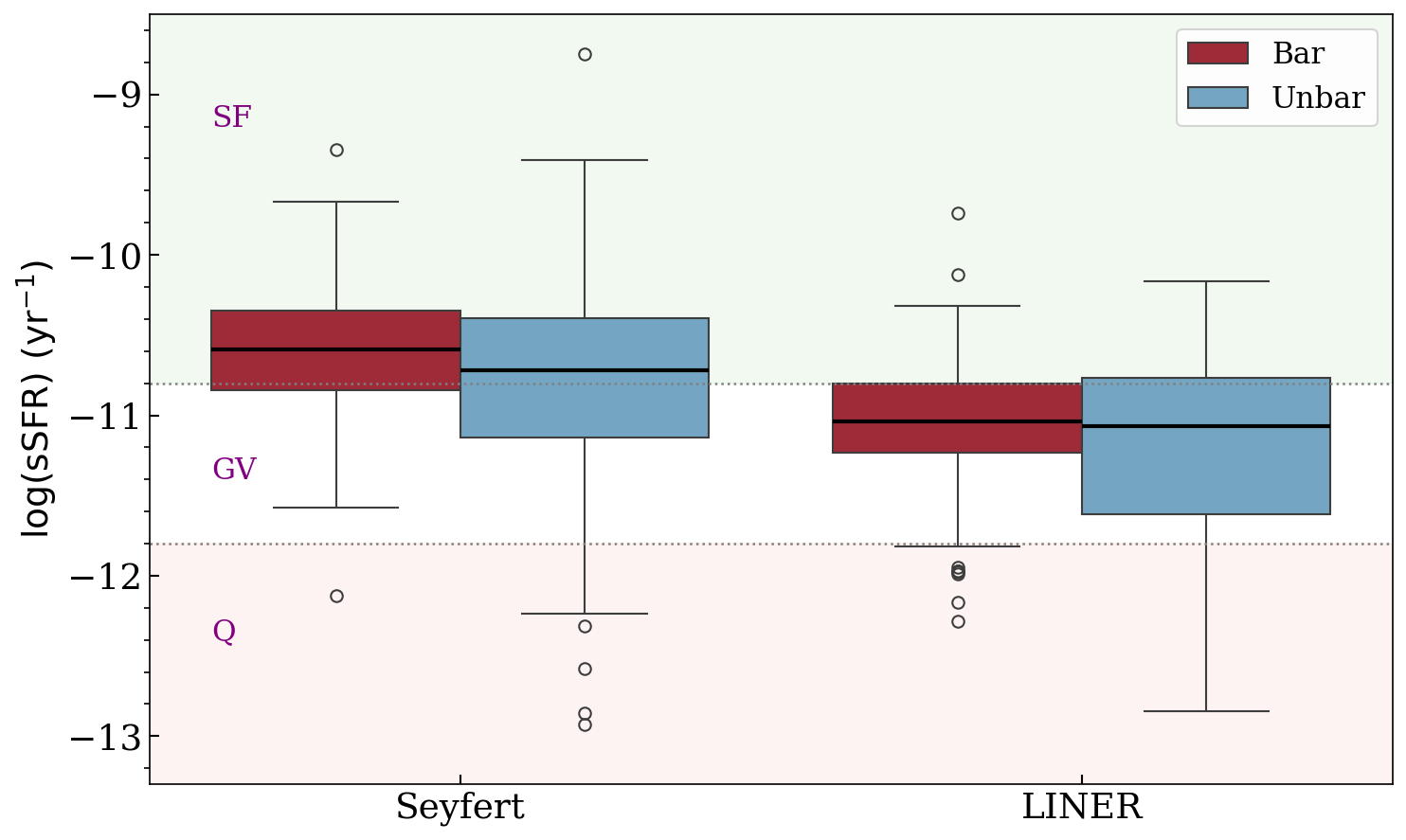}
        \caption{}
        \label{fig:sSFR_scatter_AGN}
    \end{subfigure}
    \begin{subfigure}{0.46\textwidth}
        \includegraphics[width=\textwidth]{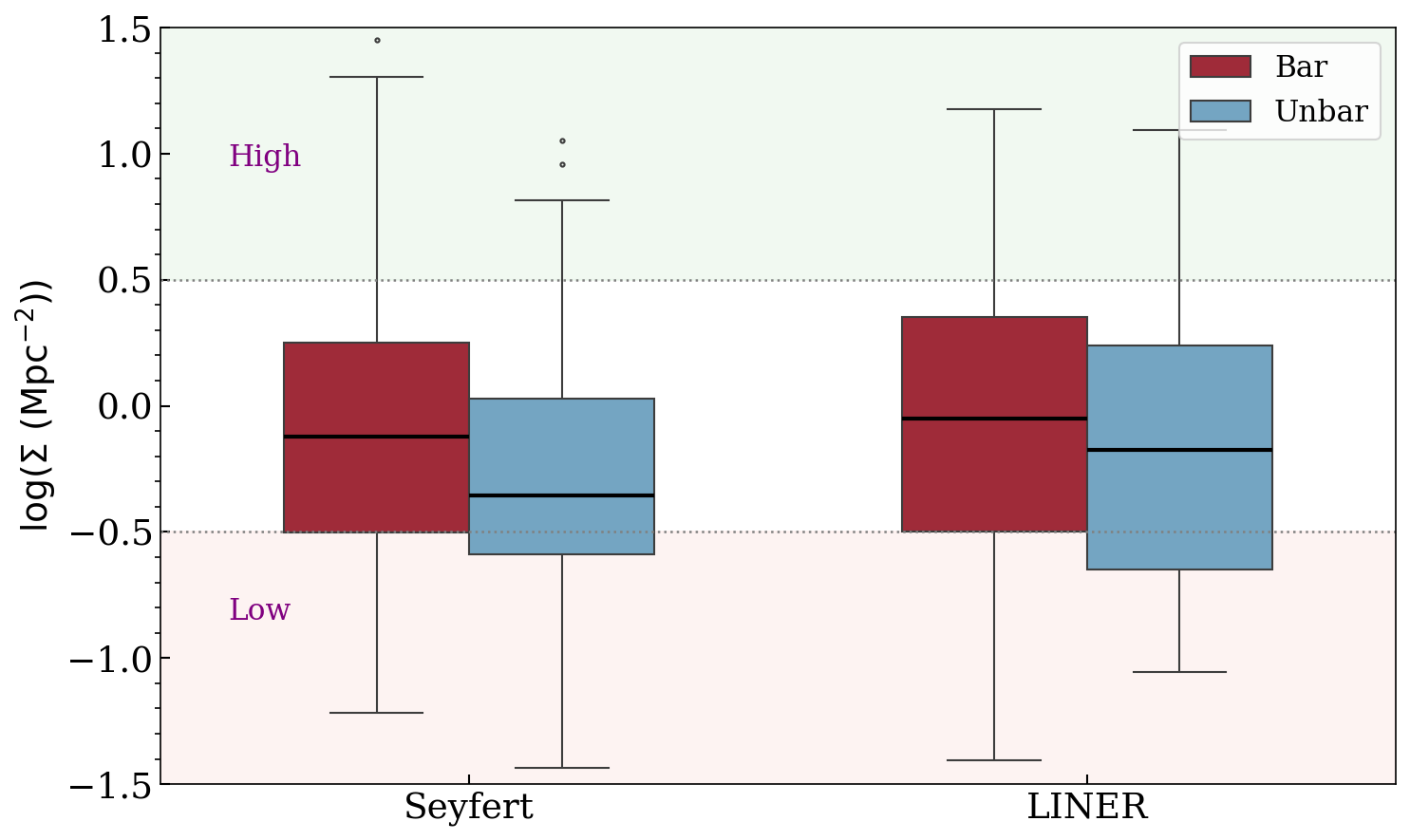}
        \caption{}
        \label{fig:density_histogram_AGN_plot}
    \end{subfigure}
    \caption{(a) Box plot distribution of barred (green) and unbarred (red) Seyferts and LINERs. \textit{Left:} sSFR distribution across SF, GV, and Q regions. \textit{Right:} Stellar mass distribution divided into low- and high-mass regimes. Open circles indicate outliers. (b) Fraction of Seyfert (magenta) and LINER (gray) as a function of local environmental density for barred (left) and unbarred (right) galaxies. Barred AGNs show a rising fraction with increasing density, while unbarred AGNs remain nearly constant. Faded points indicate bins with fewer than five galaxies.}
    \label{fig:AGN_combined}
\end{figure*}

\subsection{AGN-hosting galaxies} \label{sec:agn}
AGNs also play a crucial role in shaping galaxy evolution, as their feedback can either suppress or trigger star formation. Their influence is closely connected to galaxy morphology and environmental density, offering valuable insights into how star formation varies across different structural and environmental conditions.

\subsubsection{Star-formation}
Cross-matching these AGNs with our master sample of 7,408 galaxies yielded a final sample of 296 AGNs, consisting of 161 Seyferts and 135 LINERs. Further classification based on the presence of bars in AGN-hosting galaxies reveals an almost equal distribution, with 150 barred and 146 unbarred AGNs. Among the barred AGNs, Seyferts (88) are slightly more numerous than LINERs (62), whereas the unbarred AGN sample contains equal numbers of Seyferts (73) and LINERs (73).

When classifying AGNs by galaxy mass, only 22 out of 296 are identified as low-mass galaxies  ($\log(M_\star/M_\odot) < 10$), which are insufficient for statistically meaningful analysis. Therefore, we used the entire AGN sample without applying mass-based distinctions. 

Figure~\ref{fig:sSFR_scatter_AGN} presents the box plot distribution of the sSFR for AGNs with or without bars. The colored boxes represent the interquartile range (IQR; 25–75\%), the horizontal line marks the median value, and the whiskers indicate the full range of the non-outlier data. Open circles denote outliers in the distributions. Our results indicate that AGNs are more frequently located in the star-forming regions or near the boundary between the star-forming and green valley regimes. \citet{2023A&A...674A..85A} also reported that AGN host galaxies generally exhibit sSFRs below those of galaxies on the star-forming main sequence. In our sample, when we check for Seyferts and LINERs separately, it shows that Seyferts, particularly barred Seyferts, account for the higher incidence in star-forming regions, whereas the green valley region is primarily populated by barred LINER galaxies, as shown in Figure~\ref{fig:sSFR_scatter_AGN}.

\begin{table*}[htbp]
\centering
\caption{Regression coefficients ($\beta_{\rm Bar}$) for the bar variable in the multivariate linear regression models. Values are given as $\beta \pm$ standard error for star-forming (SF), green valley (GV), and quenched (Q) galaxies.}
\label{tab:regression_bar}
\begin{tabular}{lcccc}
\toprule
Population & Model 1 & Model 2 & Model 3 & Model 4 \\
\midrule
SF &
$0.069 \pm 0.010$ &
$0.070 \pm 0.010$ &
$0.060 \pm 0.010$ &
$0.061 \pm 0.010$ \\

GV &
$0.043 \pm 0.015$ &
$0.039 \pm 0.015$ &
$0.026 \pm 0.015^{*}$ &
$0.023 \pm 0.015^{*}$ \\

Q &
$-0.045 \pm 0.027^{*}$ &
$-0.046 \pm 0.027^{*}$ &
$-0.042 \pm 0.027^{*}$ &
$-0.043 \pm 0.027^{*}$ \\
\bottomrule
\end{tabular}

\vspace{2mm}
\begin{minipage}{0.95\linewidth}
\footnotesize
\textbf{Notes.}
Model~1 includes stellar mass, local density, and bar.
Model~2 additionally includes AGN.
Model~3 includes stellar mass, local density, bar, and morphology.
Model~4 includes stellar mass, local density, bar, AGN, and morphology.
In all these models, bars, AGN, and morphology are added as categorical variables. Non-significant (p$>$0.05) values are denoted with $\star$.
\end{minipage}

\end{table*}

A closer examination of these statistics reveals that barred Seyfert galaxies have median values located near the boundary between the star-forming and green valley regions, with a stronger concentration in the star-forming regime. Their unbarred counterparts, however, display a wider distribution across all three regions, with a median sSFR still within the star-forming range. This trend, illustrated in the Figure~\ref{fig:sSFR_scatter_AGN}, suggests that the presence of a bar helps maintain Seyfert galaxies near the transition between the star-forming and green valley regions. In contrast, both barred and unbarred LINER galaxies show similar median positions centered around the green valley region. However, unbarred LINERs exhibit a broader spread toward both the star-forming and quenched regions. 

This analysis suggests that bars in Seyfert galaxies help maintain their position near the star-forming–green valley transition, potentially aiding in the sustained formation of stars. In contrast, bars in LINER galaxies appear to confine them within the green valley region, facilitating their progression toward quiescence. Previous studies have also shown that AGN hosts are frequently associated with ongoing or enhanced central star formation \citep{2011MNRAS.418.2043E, 2013ApJ...771...63R}, and that bars can trigger AGN activity \citep{2020MNRAS.494.5839K, 2024MNRAS.532.2320G, 2025A&A...699A.204M}, supporting our finding that bars in Seyfert galaxies tend to keep them near the star-forming–green valley transition.

\subsubsection{Effect of local environmental}
To further investigate this, it is important to examine whether the environment influences the distribution of galaxies hosting AGN. For this purpose, we analyzed the AGN-wise distribution of galaxies across different environmental regions, as shown in Figure~\ref{fig:density_histogram_AGN_plot}.

Barred AGNs exhibit a higher median value within dense environments, suggesting that dense environments may enhance bar-driven AGN fueling. In contrast, unbarred AGNs prefer less dense environments. This trend suggests that bars play a more significant role in triggering AGN activity under external environmental influences, whereas unbarred galaxies likely rely more on internal processes.

Previous studies \citep{2014A&A...572A..86A, 2020ApJ...901L..38K} have also emphasized the role of dense environments in triggering AGN fueling in barred galaxies. Our analysis suggests that bars support star formation in Seyferts and promote transition towards quenching in LINERs, with their influence modulated by the surrounding environment.

\subsection{Multivariate regression analysis} \label{sec:multi}
The preceding sections show that the star formation properties of barred and unbarred galaxies vary with stellar mass, morphology, and local environment. Since these galaxy properties are mutually correlated, the observed differences in sSFR may reflect the combined influence of multiple factors. We therefore perform a multivariate linear regression analysis to assess the association between stellar bars and sSFR while simultaneously accounting for stellar mass, local environmental density, AGN activity, and galaxy morphology.

The dependent variable is the
$\log({\rm sSFR})$, and the regression model is written as $\log({\rm sSFR}) = \beta_0 +\beta_1\log(M_\star) +\beta_2\log(\Sigma) +\beta_3{\rm (Bar)} +\beta_4{\rm (AGN)} +\beta_5{\rm (Morphology)}$, where $\beta$'s are the coefficients of corresponding quantities.

The variables Bar, AGN, and Morphology are treated as categorical variables. Four regression models are constructed to examine the robustness of the bar coefficient. Model~1 includes stellar mass, local density, and bar. Model~2 additionally includes AGN activity. Model~3 includes morphology while excluding AGN activity. Model~4 includes stellar mass, local density, bar, AGN activity, and morphology. The regression analysis is performed using the \texttt{statsmodels} package in Python \citep{seabold2010}. 

Table~2 summarizes the regression coefficient of the bar term for the four models. For star-forming galaxies, the bar coefficient remains positive in all four models, with values ranging from 0.060 to 0.070. The coefficient decreases slightly after morphology is included in the regression, but remains statistically significant, indicating that the positive association between bars and sSFR persists after accounting for stellar mass, local density, AGN activity, and morphology.

For green valley galaxies, the bar coefficient is positive in Models~1 and 2, but decreases after morphology is included and becomes statistically insignificant in Models~3 and 4. This suggests that the observed difference in sSFR between barred and unbarred green valley galaxies is not statistically significant after accounting for morphology.

For quenched galaxies, the bar coefficient is negative in all four models and remains statistically insignificant. This indicates that no significant association between stellar bars and sSFR is found in quenched galaxies after accounting for the other galaxy properties included in the regression.

The regression results indicate that the bar--sSFR association is strongest for star-forming galaxies and becomes less evident towards other star-formation states. The persistence of the positive bar coefficient for star-forming galaxies across all four models suggests that this association remains after accounting for the other galaxy properties considered in the analysis.

\begin{figure*}[htbp]
    \centering
    \begin{subfigure}[t]{0.32\textwidth}
        \includegraphics[width=\textwidth]{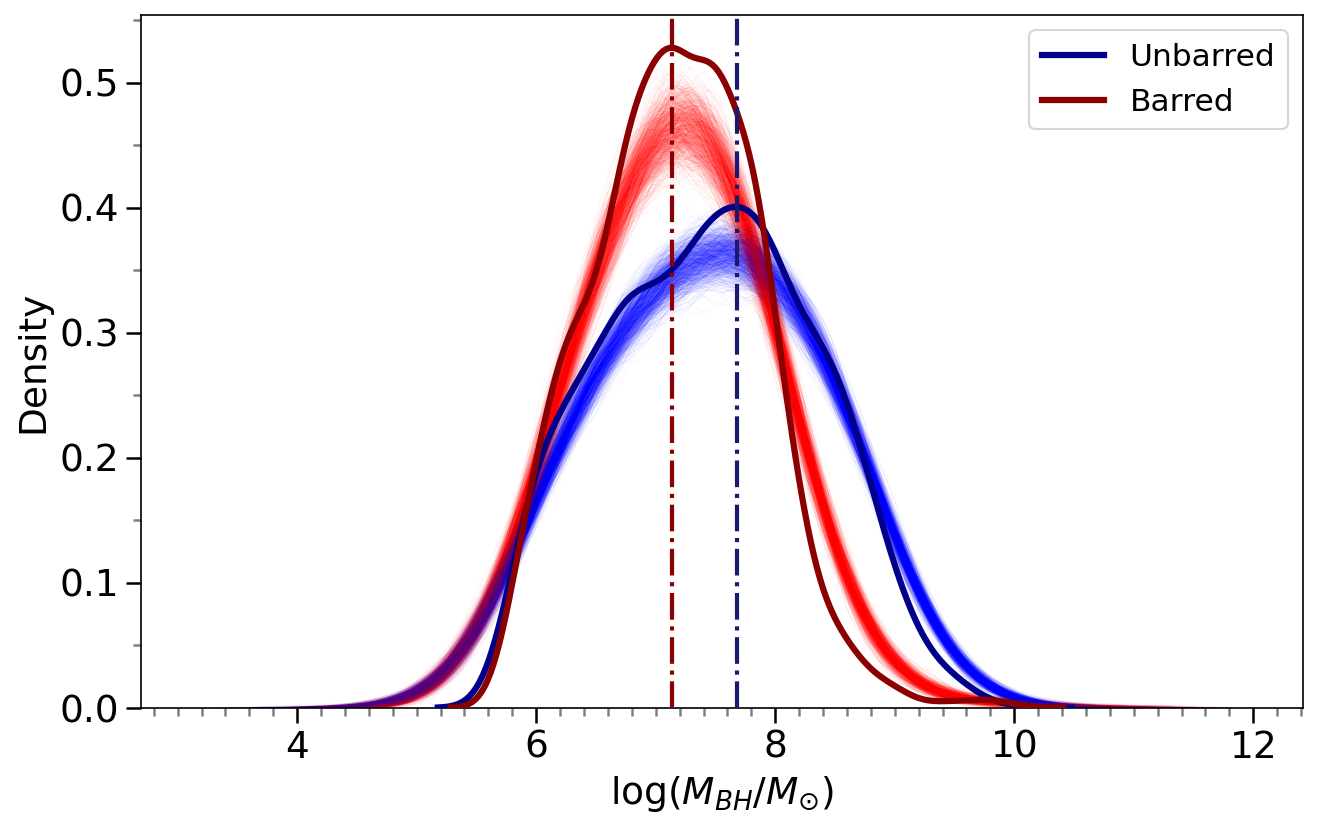}
        \caption{\scriptsize }
        \label{fig:bar_unbar_mbh}
    \end{subfigure}
    \begin{subfigure}[t]{0.32\textwidth}
        \includegraphics[width=\textwidth]{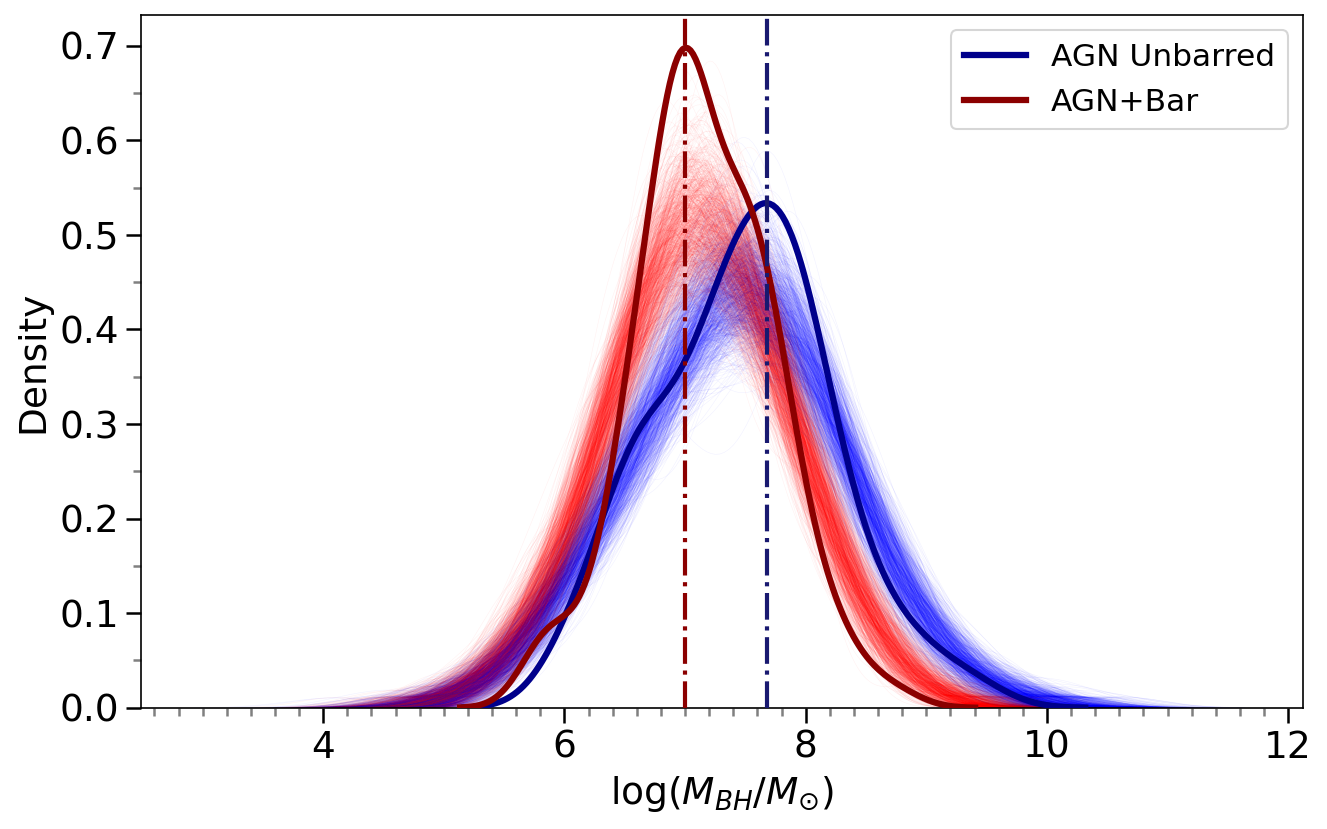}
        \caption{\scriptsize}
        \label{fig:agn_mbh_bar_unbar}
    \end{subfigure}
    \begin{subfigure}[t]{0.32\textwidth}
        \includegraphics[width=\textwidth]{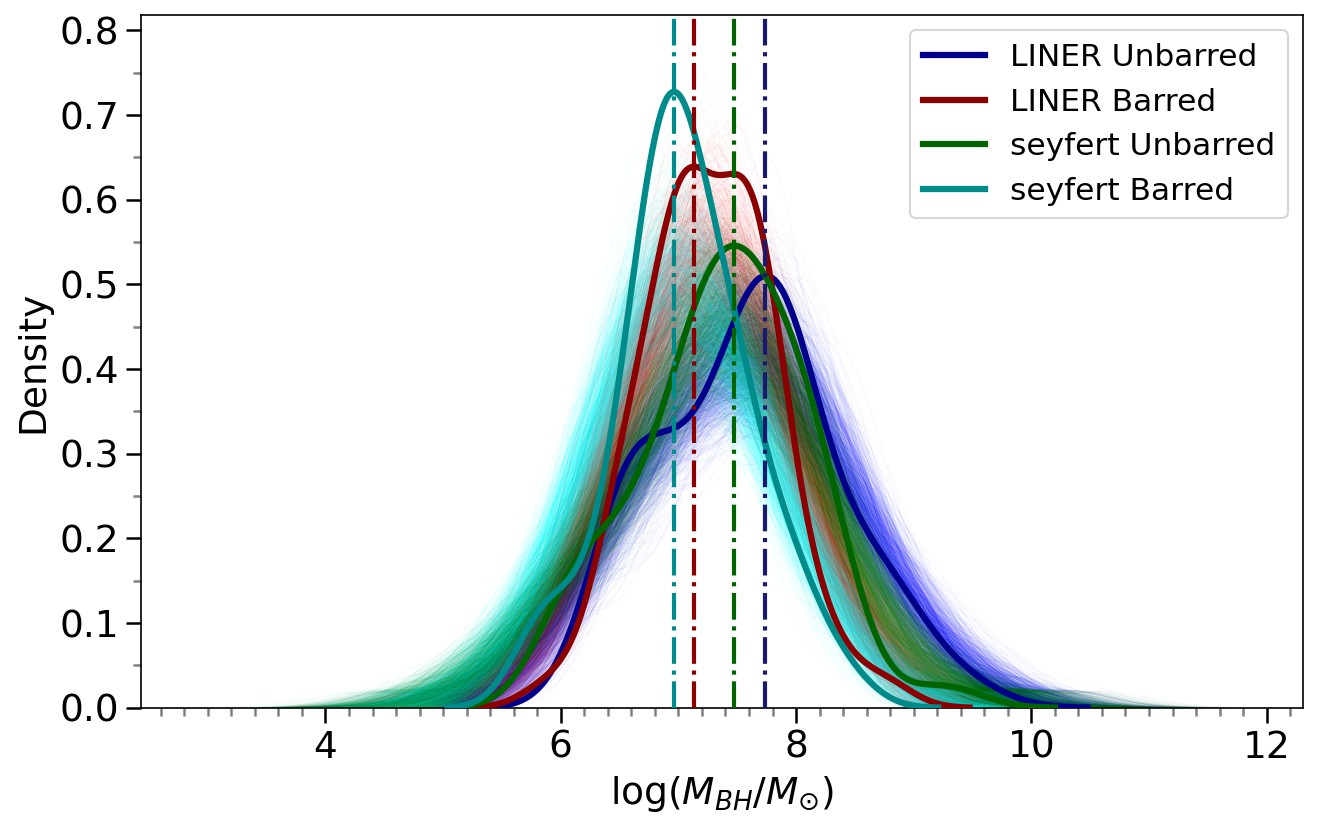}
        \caption{\scriptsize}
        \label{fig:liner_seyfert_mbh_bar_unbar}
    \end{subfigure}\\[1ex]
    \begin{subfigure}[t]{0.32\textwidth}
        \includegraphics[width=\textwidth]{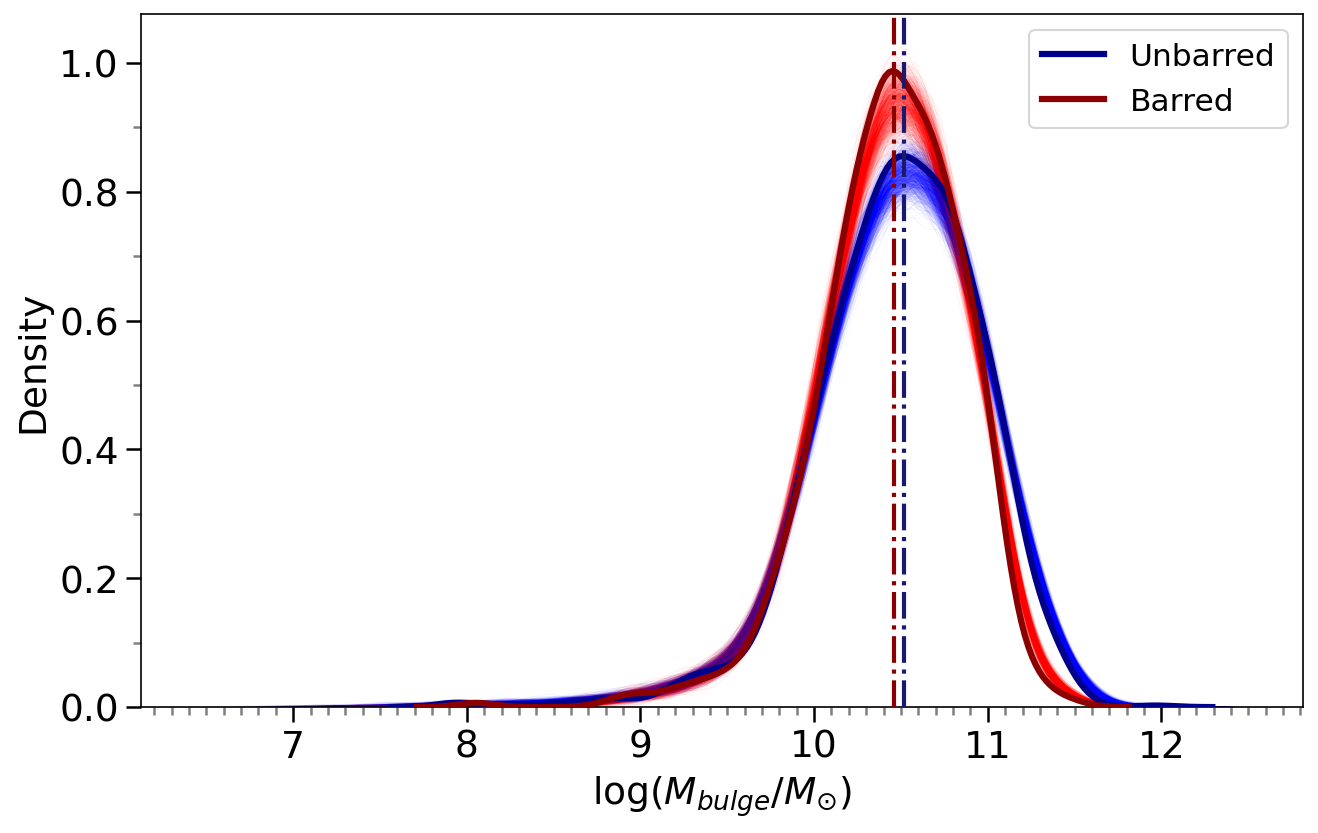}
         \caption{\scriptsize}
        \label{fig:bar_unbar_mbulge}
    \end{subfigure}
    \begin{subfigure}[t]{0.32\textwidth}
        \includegraphics[width=\textwidth]{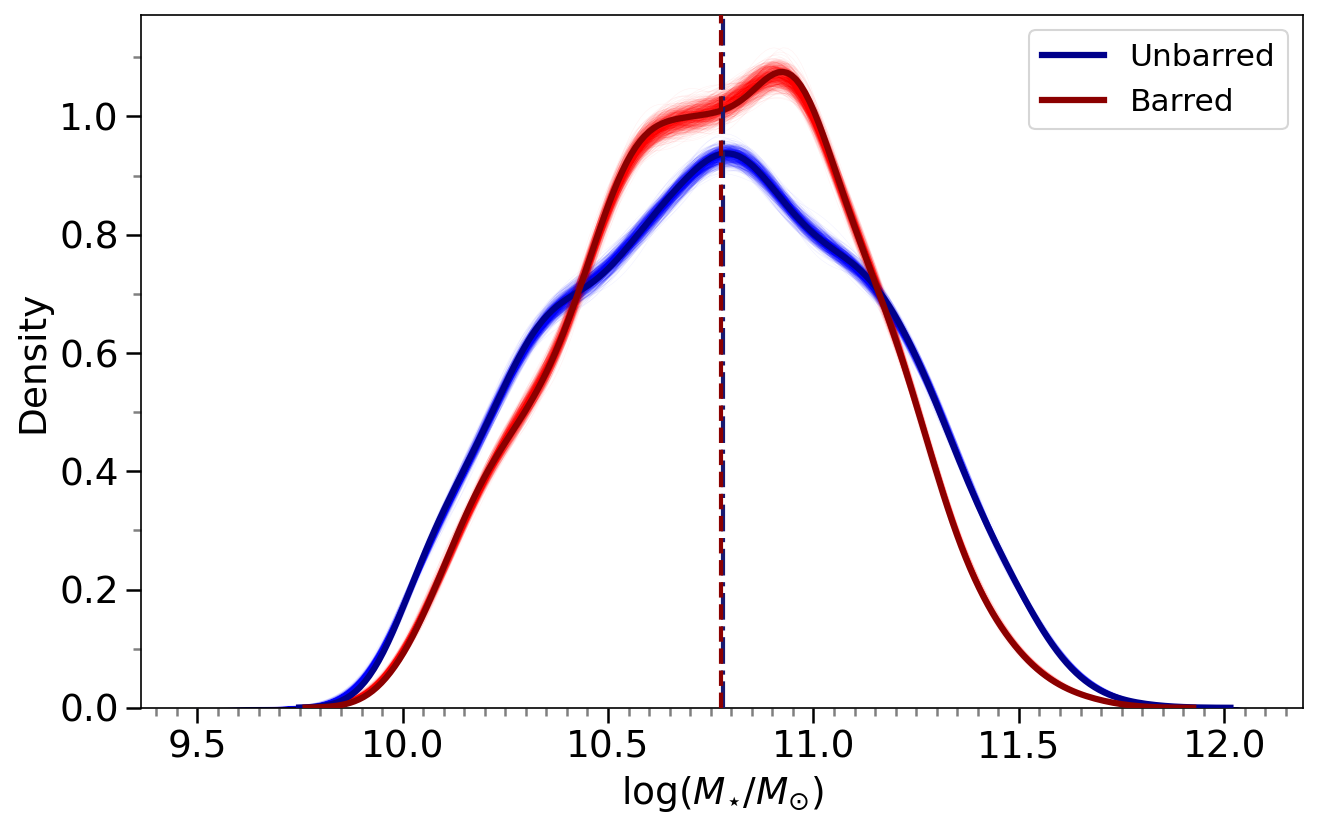}
         \caption{\scriptsize}
        \label{fig:bar_unbar_mstar}
    \end{subfigure}
    \begin{subfigure}[t]{0.32\textwidth}
        \includegraphics[width=\textwidth]{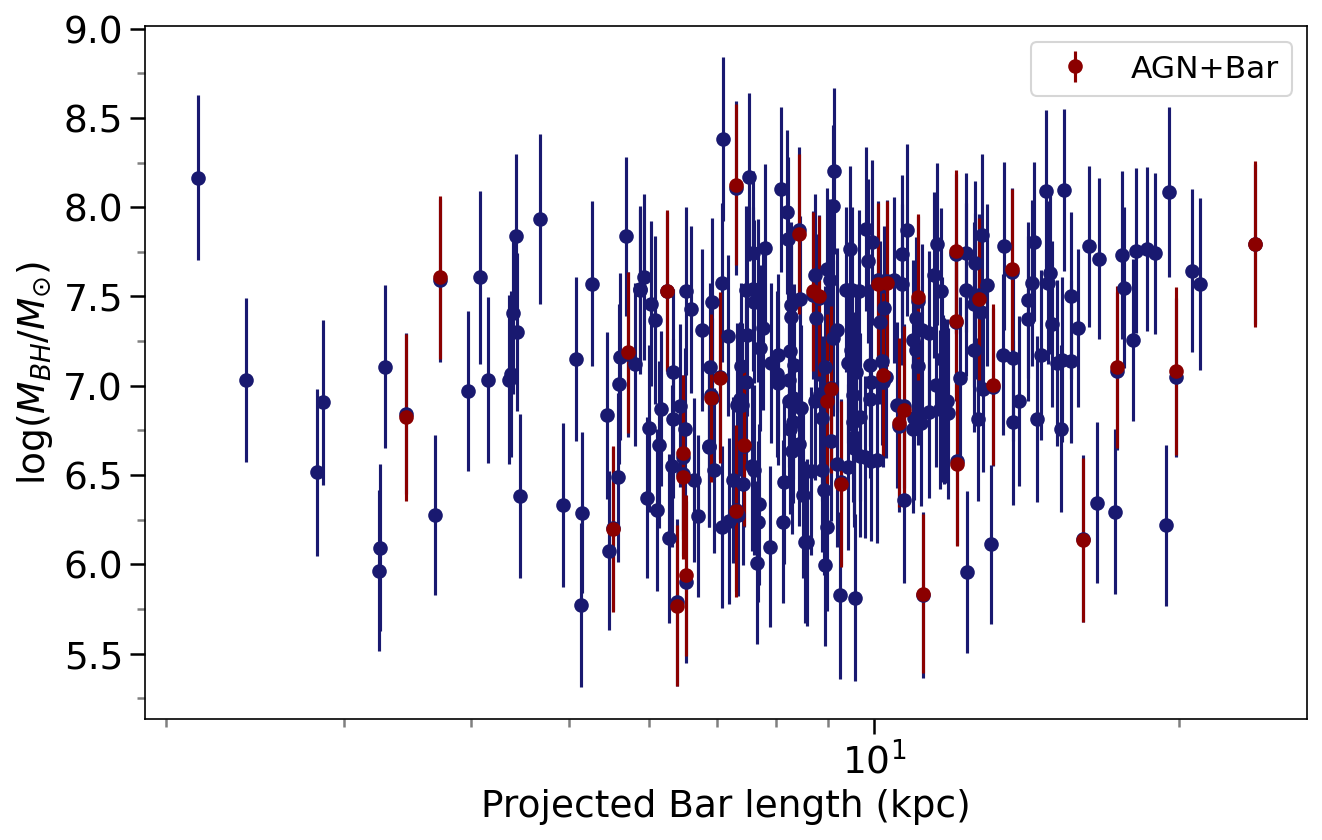}
        \caption{\scriptsize}
        \label{fig:bar_length_mbh}
    \end{subfigure}
    \caption{Kernel density estimates (KDEs) of the distributions of (a) black hole mass, (b) bulge mass, and (c) stellar mass for barred and unbarred galaxies. Panel (d) shows the black hole mass distribution for the AGN subset, while panel (e) presents the corresponding distributions for the LINER and Seyfert subclasses. Solid curves represent the observed distributions, and the dashed vertical lines indicate the median values. The shaded regions in panels (a), (d), and (e) correspond to 1,000 mock realizations generated by perturbing the black hole masses with Gaussian noise using intrinsic scatters of 0.34 dex for S0 galaxies and 0.46 dex for ETS and LTS galaxies. For panels (b) and (c), the mock realizations are generated using the individual measurement uncertainties of the bulge and stellar masses, respectively. Panel (f) shows the distribution of projected bar length and black hole mass for the sample of 410 barred galaxies (blue), with AGN-hosting barred galaxies highlighted in maroon.}
    \label{fig:bar_ubar_trends}
\end{figure*}

\subsection{Relation with Black Hole Growth}
We further explore the possible role of stellar bars in the growth of central black holes. For this analysis, we estimate black hole masses for disk galaxies with stellar mass $M_\star > 10^{10}\,M_\odot$, including S0s, ETS, and LTS. The stellar mass cut is applied to focus on massive disk galaxies where central bulges are well developed and the $M_{\rm BH}-\sigma$ relation is known to be more robustly established \citep{2013ARA&A..51..511K}. In addition, we impose a velocity dispersion threshold of $\sigma > 70$ km s$^{-1}$ to avoid measurements close to the instrumental resolution limit of the MaNGA survey. These selection criteria ensure reliable velocity dispersion measurements and allow a meaningful comparison of the black hole mass distributions between barred and unbarred galaxies, enabling us to investigate whether the presence of a bar is associated with enhanced black hole growth.

Black hole masses are estimated using the $M_{\rm BH}-\sigma$ relation from \citet{2013ApJ...764..184M}. We adopt the relation calibrated for early-type galaxies (S0s) and late-type galaxies (spirals: ETS and LTS). The stellar velocity dispersion ($\sigma$) is taken from the Pipe3D catalog, using measurements from the central region (2.5$^{\prime\prime}$) of each galaxy. With these criteria, we have a sample of 3038 galaxies, including 1315 barred and 1723 unbarred galaxies. Among these, 228 galaxies host AGN, of which 126 are barred, and 102 are unbarred.

To evaluate the statistical significance of the observed differences between barred and unbarred galaxies, we perform a Monte Carlo approach that accounts for the intrinsic scatter in the $M_{\rm BH}-\sigma$ relation ($\epsilon = 0.34$ dex for S0s and $\epsilon = 0.46$ dex for spiral galaxies; \citet{2013ApJ...764..184M}). For each galaxy, we generate 1,000 simulated black hole mass estimates by adding random Gaussian noise scaled to the morphology-dependent intrinsic scatter. Figure~\ref{fig:bar_ubar_trends} shows a comparison between the peaks of the simulated kernel density distributions and the original observed peaks. This probabilistic approach enables us to test the robustness of the mass differences between barred and unbarred populations despite the intrinsic uncertainties in the scaling relations.

We find that barred galaxies exhibit systematically lower black hole masses compared to unbarred galaxies, with an average difference of $0.30 \pm 0.03$ dex (Figure~\ref{fig:bar_unbar_mbh}). This trend becomes more pronounced when considering only AGN hosts, where the offset increases to $0.42 \pm 0.11$ dex (Figure~\ref{fig:agn_mbh_bar_unbar}). Within the AGN population, Seyfert galaxies show a difference of $0.38 \pm 0.14$ dex, while LINERs show a larger offset of $0.49 \pm 0.18$ dex (Figure~\ref{fig:liner_seyfert_mbh_bar_unbar}), with barred galaxies consistently hosting lower black hole masses than their unbarred counterparts. These results suggest that the presence of a bar does not necessarily lead to enhanced black hole growth.

To assess the statistical significance of the difference between barred and unbarred galaxies, we performed a Kolmogorov–Smirnov (KS) test, which evaluates whether two samples are drawn from the same parent distribution. The KS statistic D, which represents the maximum separation between the cumulative distributions of the two samples, yields D=0.310 (p=6.8$\times$10$^{-3}$) for LINERs, D=0.293 (p=1.1$\times$10$^{-2}$) for Seyferts, D=0.268 (p=4.4$\times$10$^{-4}$) for AGN hosts, and D=0.195 (p=3.5$\times$10$^{-25}$) for the full galaxy sample, confirming that the distributions are statistically different.

In this analysis, the stellar velocity dispersion traces the gravitational potential of the bulge. Therefore, the observed difference in black hole mass between barred and unbarred galaxies could also reflect differences in the stellar or bulge masses of their host galaxies. To test this possibility, we compared the stellar and bulge mass distributions of the two samples using kernel density estimates (Figure~\ref{fig:bar_unbar_mbulge} and ~\ref{fig:bar_unbar_mstar}). We cross-matched our sample with the bulge--disk decomposition catalog given by \citet{2014ApJS..210....3M}, obtaining bulge mass estimates for 1,134 barred galaxies and 1,512 unbarred galaxies. The median offsets between the barred and unbarred samples are $0.049 \pm 0.010$ dex in bulge mass and $0.001 \pm 0.004$ dex in stellar mass. The KS test yields $D=0.08$ ($p=8.48\times10^{-4}$) for the bulge mass distribution and $D=0.05$ ($p=2.69\times10^{-2}$) for the stellar mass distribution. Although these differences are statistically detectable given the large sample size, their magnitudes are small.

The stellar and bulge mass distributions of barred and unbarred galaxies are broadly similar. In particular, the negligible offset in stellar mass and the small offset in bulge mass are substantially smaller than the difference observed in black hole mass. This suggests that the lower black hole masses inferred for barred galaxies are unlikely to arise solely from systematic differences in the stellar or bulge masses of their host galaxies.

To further investigate this, we cross-matched our sample with the catalogs of bar lengths from \citet{2011MNRAS.415.3627H} and \citet{Geron_2023}, obtaining bar lengths for 288 and 102 galaxies, respectively. Using a sample of 410 galaxies, we examined the relation between projected bar length and black hole mass and found no evidence of any correlation, even when restricting the sample to AGN hosts (Figure~\ref{fig:bar_length_mbh}). The distribution appears completely scattered, indicating that bar length does not significantly affect black hole mass growth.

\begin{figure}
    \centering
    \includegraphics[width=\columnwidth]{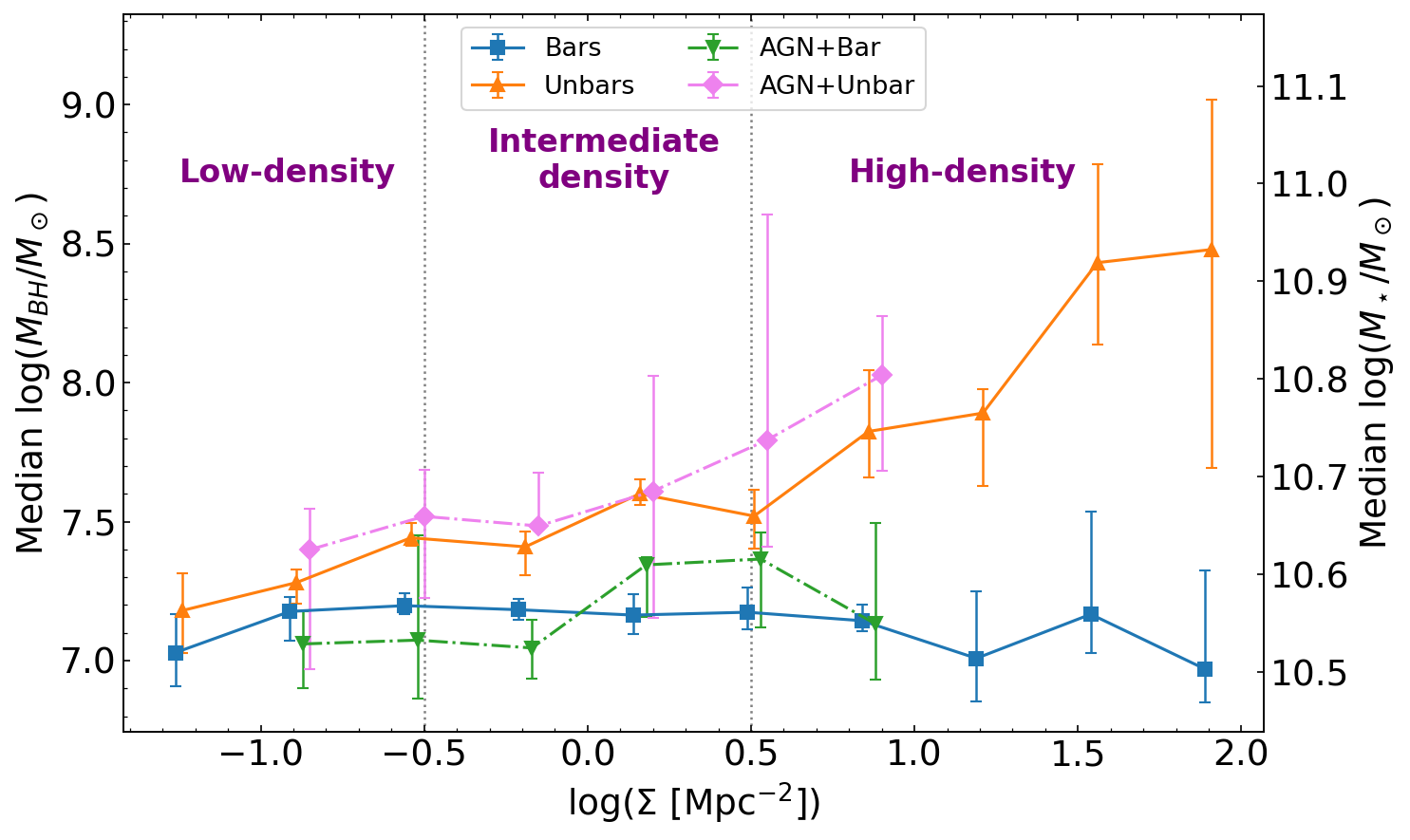}
    \caption{Median black hole mass as a function of local environmental density for barred (blue), unbarred (orange), barred AGN (green), and unbarred AGN (violet) galaxies. The vertical dotted lines divide the low-, intermediate-, and high-density environments.}
    \label{fig:mbh_env}
\end{figure}

Finally, we explore the environmental dependence of black hole mass in barred and unbarred galaxies (Figure~\ref{fig:mbh_env}). We find that in low-density environments, the difference in black hole mass between barred and unbarred galaxies is relatively small. However, the difference becomes more pronounced toward higher-density environments, where barred galaxies tend to host systematically lower black hole masses than unbarred galaxies, even among AGN hosts. This suggests that the large-scale environment plays a more important role in regulating black hole growth than the presence of a stellar bar.

Our results can be placed in the broader context of previous studies investigating the connection between stellar bars and black hole growth or AGN activity. Several studies have reported a weak or absent direct connection between bars and AGN triggering. \citet{Lee2012} found no clear evidence that stellar bars play a dominant role in triggering AGN activity. Similarly, \citet{SilvaLima2022} showed that although AGN are more frequently found in barred galaxies and may exhibit higher accretion rates, the dependence strongly relies on the adopted $M_{\rm BH}-\sigma$ relation, and no correlation was found between AGN activity and bar strength, suggesting that additional mechanisms may also contribute to fueling the central black hole. On the other hand, some studies have suggested a positive link between bars and black hole growth. For instance, \citet{Kataria2024} reported that barred galaxies tend to host larger black hole masses and higher average accretion rates since the epoch of bar formation, indicating that stellar mass and long-term bar-driven inflow may influence black hole growth, with bar-driven growth being independent of the environment. Also, more recently, \citet{2025A&A...699A.204M} found that barred galaxies tend to host smaller black holes, but also noted enhanced AGN activity in galaxies with strong bars in both low- and high-density environments. In comparison, our results show that barred galaxies systematically host lower black hole masses than unbarred galaxies, and we find no correlation between bar length and black hole mass. Together with the environmental trends observed in our sample, this suggests that while bars may redistribute gas within galaxies, their role in directly driving black hole growth is likely secondary compared to other factors such as galaxy environment.

We note that the observed offset in black hole mass between barred and unbarred galaxies may be influenced by differences in stellar mass and morphology distributions between the two populations. A detailed control of these factors, including matched samples or multivariate analysis, is beyond the scope of the present work. Instead, our study aims to highlight the observed statistical trends within the MaNGA sample and place them in the broader context of the connection between stellar bars and central black holes.

The results presented in this work provide an empirical baseline for interpreting the role of internal structures in galaxy evolution, particularly in the context of high-redshift galaxies now being observed with facilities such as JWST, where the interplay between stellar bars, star formation, and black hole growth remains an open question.

\section{Conclusions and Summary} \label{sec:conclusion}
We analyze a sample of 7,408 MaNGA galaxies to examine statistical associations between star formation, stellar bars, host galaxy morphology, and local environmental density across a broad stellar mass range. This sample consists of 2,641 barred galaxies. Rather than separating the sample only into low- and high-mass, we examine the barred and unbarred populations across different stellar mass intervals ($\leq\,9.5, 9.5-10, 10-10.5, 10.5-11, \text{and} \geq\,11 $ in $\log(M_\star/M_\odot)$) to investigate how these associations vary with stellar mass, morphology, and environment. We also examine AGN-hosting galaxies and the connection between stellar bars and black hole mass. The main findings are summarized as follows:

\begin{enumerate}

    \item Barred galaxies show higher median sSFR than unbarred galaxies across most stellar mass intervals. Considering only disk galaxies, the largest difference is found for galaxies with $\log(M_\star/M_\odot) \geq 11$, where barred galaxies show a median sSFR higher by 0.439 dex than unbarred galaxies. Across the different stellar mass intervals, the barred-unbarred difference is more pronounced at higher stellar masses.

    \item Along with stellar mass, host galaxy morphology is also associated with differences in global sSFR. LTS galaxies show positive differences in median sSFR in the intermediate-mass intervals, whereas S0 galaxies show negative differences in the same mass ranges. At the highest stellar masses, the S0 population shows a positive difference, although this result is based on a limited number of barred S0 galaxies. These results indicate that the association between bars and sSFR is not uniform across different disk morphologies.

    \item The environmental trends of barred and unbarred galaxies are broadly similar within the same stellar mass intervals. Although differences are present in some morphology–mass and star-formation–mass intervals, the overall environmental trends of barred and unbarred galaxies remain broadly comparable. This suggests that the barred–unbarred differences in sSFR cannot be attributed solely to variations in local density.

    \item The sSFR distributions of barred AGN hosts differ between Seyferts and LINERs. Barred Seyfert galaxies tend to cluster near the star-forming--green valley boundary, whereas barred LINER galaxies preferentially occupy the green valley region. The environment appears to play a secondary role, with barred AGN hosts residing in relatively denser regions compared to unbarred ones.

    \item The multivariate regression analysis shows that the bar–sSFR association depends on the star-formation state of the galaxies. For star-forming galaxies, the positive bar coefficient remains statistically significant after accounting for stellar mass, local density, AGN activity, and morphology. For green valley galaxies, the bar coefficient becomes statistically insignificant after morphology is included, while no statistically significant association is found for quenched galaxies.

    \item Our results indicate that barred galaxies host systematically lower black hole masses than unbarred galaxies by $0.30 \pm 0.03$ dex, with a larger offset of $0.42 \pm 0.11$ dex among AGN hosts. We also find that the presence and size of stellar bars do not strongly influence black hole growth. The observed differences between barred and unbarred galaxies are more consistent with environmental effects, suggesting that external processes may play a more important role in governing the evolution of central black holes.
    
\end{enumerate}

These results emphasize the interconnected roles of internal galaxy properties, such as stellar bars and AGN activity, and external environmental conditions in shaping galaxy properties. Within this broader context, stellar bars are associated with enhanced global sSFR in star-forming galaxies, with the strength of this association varying across stellar mass and morphology. The observed correlations between morphology, sSFR, and environment further highlight the importance of considering these properties together when interpreting the role of stellar bars in galaxy evolution. Our findings provide a valuable benchmark for upcoming large-scale IFU surveys such as SDSS-V, 4MOST, and HECTOR, which will extend these studies to higher redshifts and statistically larger samples.

\section*{ACKNOWLEDGMENTS}
We thank the anonymous referee for thoughtful comments and suggestions that significantly improved the presentation and focus of this work. Funding for the Sloan Digital Sky Survey IV has been provided by the Alfred P. Sloan Foundation, the U.S. Department of Energy Office of Science, and the Participating Institutions. SDSS-IV acknowledges support and resources from the Center for High-Performance Computing at the University of Utah. The SDSS web site is www.sdss.org. SDSS-IV is managed by the Astrophysical Research Consortium for the Participating Institutions of the SDSS Collaboration including the Brazilian Participation Group, the Carnegie Institution for Science, Carnegie Mellon University, the Chilean Participation Group, the French Participation Group, Harvard-Smithsonian Center for Astrophysics, Instituto de Astrof\'isica de Canarias, The Johns Hopkins University, Kavli Institute for the Physics and Mathematics of the Universe (IPMU) / University of Tokyo, Lawrence Berkeley National Laboratory, Leibniz Institut f\"ur Astrophysik Potsdam (AIP), Max-Planck-Institut f\"ur Astronomie (MPIA Heidelberg), Max-Planck-Institut f\"ur Astrophysik (MPA Garching), Max-Planck-Institut f\"ur Extraterrestrische Physik (MPE), National Astronomical Observatories of China, New Mexico State University, New York University, University of Notre Dame, Observat\'ario Nacional / MCTI, The Ohio State University, Pennsylvania State University, Shanghai Astronomical Observatory, United Kingdom Participation Group, Universidad Nacional Aut\'onoma de M\'exico, University of Arizona, University of Colorado Boulder, University of Oxford, University of Portsmouth, University of Utah, University of Virginia, University of Washington, University of Wisconsin, Vanderbilt University, and Yale University. \\

\textit{Softwares}: Topcat \citep{2005ASPC..347...29T}, Matplotlib \citep{Hunter:2007}, NumPy  \citep{harris2020array}, Seaborn \citep{Waskom2021} 

\appendix
\renewcommand{\thetable}{\Alph{section}\arabic{table}}
\renewcommand{\thefigure}{\Alph{section}\arabic{figure}}
\section{Additional Tables and Figures} \label{appendix:table}
\setcounter{table}{0}
\setcounter{figure}{0}

\begin{figure*}[htbp]
    \centering
    \begin{subfigure}[t]{0.395\textwidth}
        \centering
        \includegraphics[width=\linewidth]{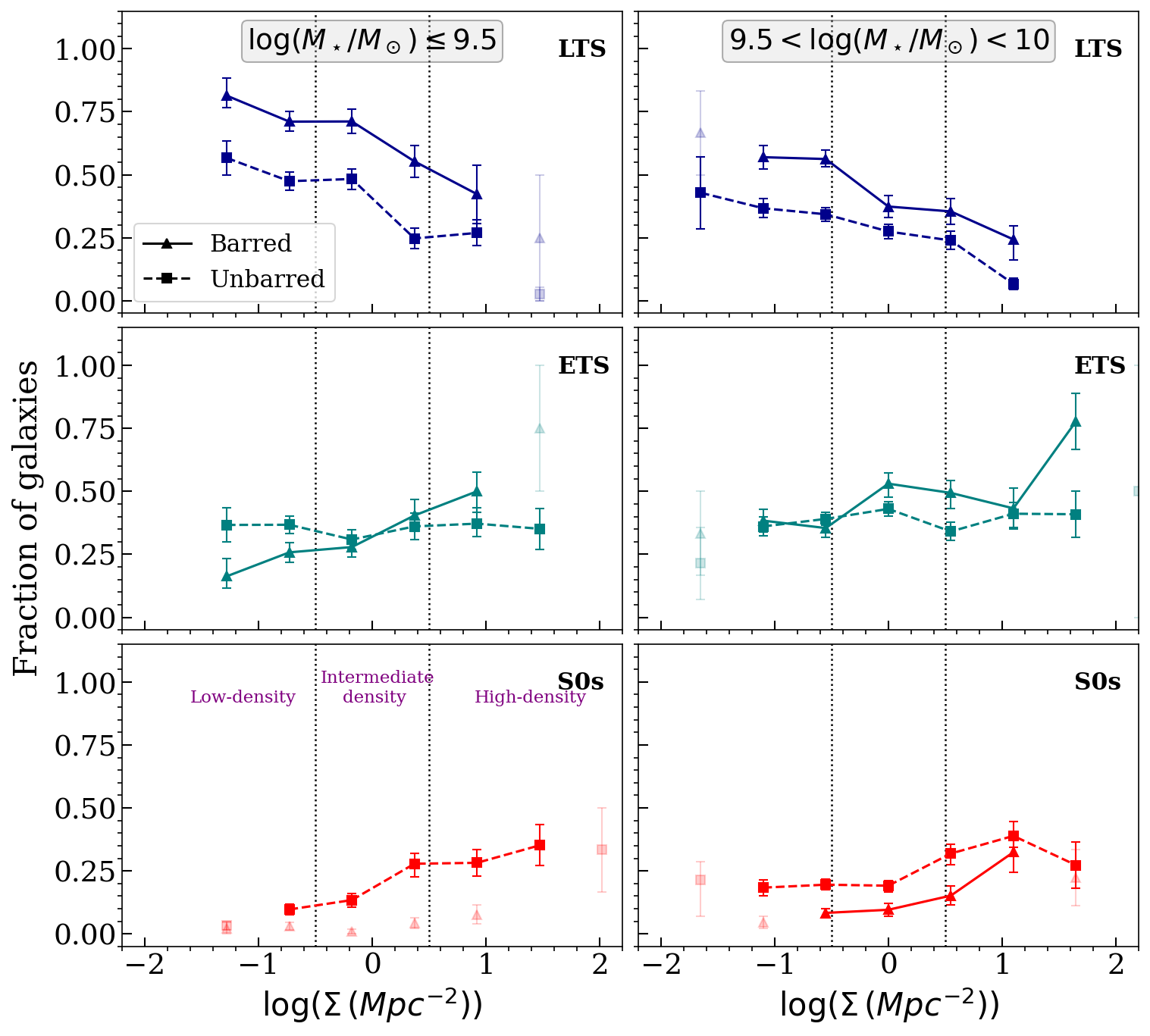}
    \end{subfigure}\hspace{-1.7mm}
    \begin{subfigure}[t]{0.59\textwidth}
        \centering
        \includegraphics[width=\linewidth]{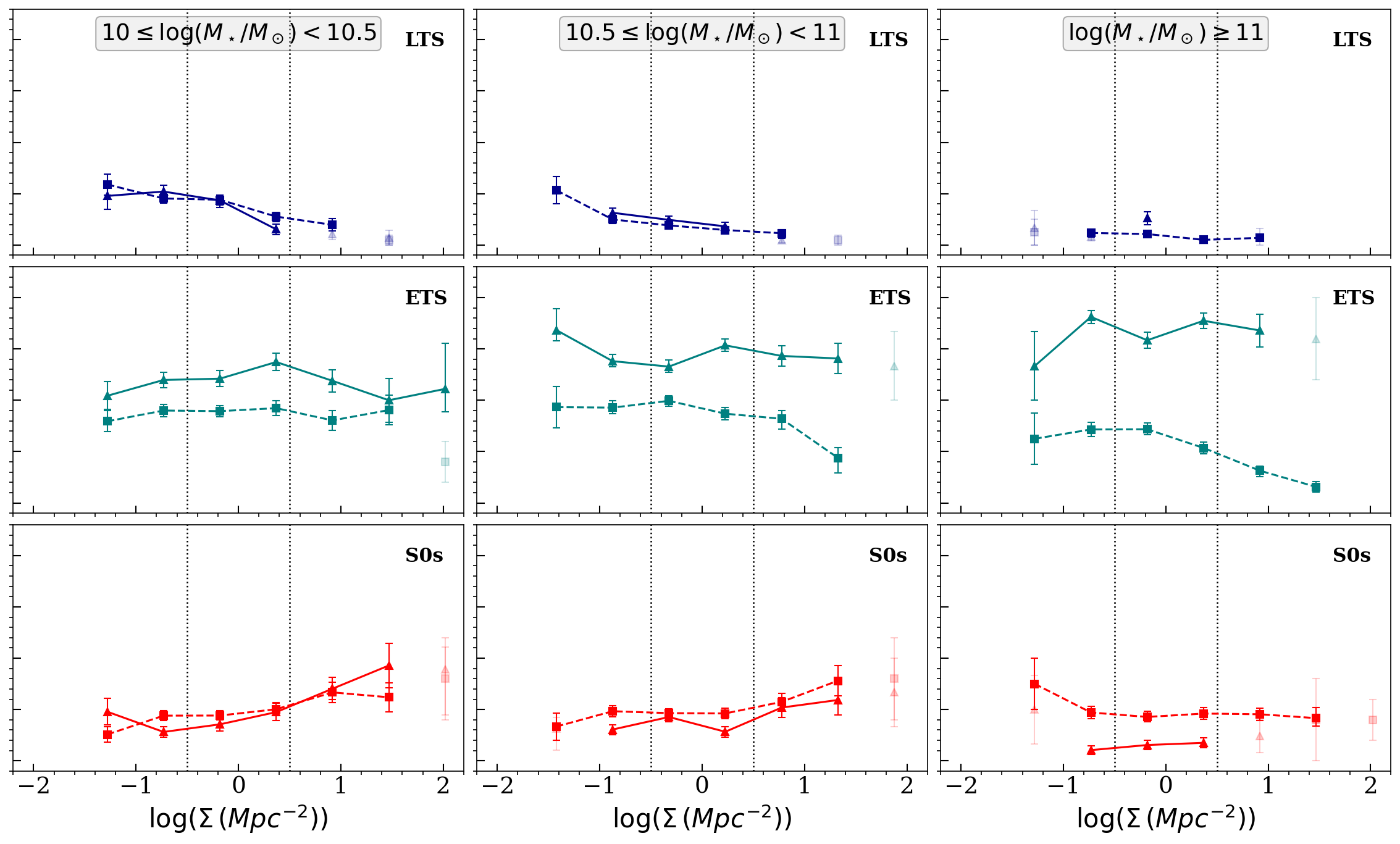}
    \end{subfigure}
    \caption{Distributions of the morphological fractions of barred (triangle) and unbarred (square) galaxies for different stellar mass bins as a function of local environmental density. The rows correspond to LTS (blue), ETS (teal), and S0s (red). The vertical dotted lines divide the low-, intermediate-, and high-density environments. Bin size = 0.55 Mpc$^{-2}$. Faded points indicate bins with fewer than five galaxies. Error bars are estimated using bootstrap resampling.}
    \label{fig:bar_morph}
\end{figure*}

\begin{figure*}[htbp]
    \centering
    \begin{subfigure}[t]{0.395\textwidth}
        \centering
        \includegraphics[width=\linewidth]{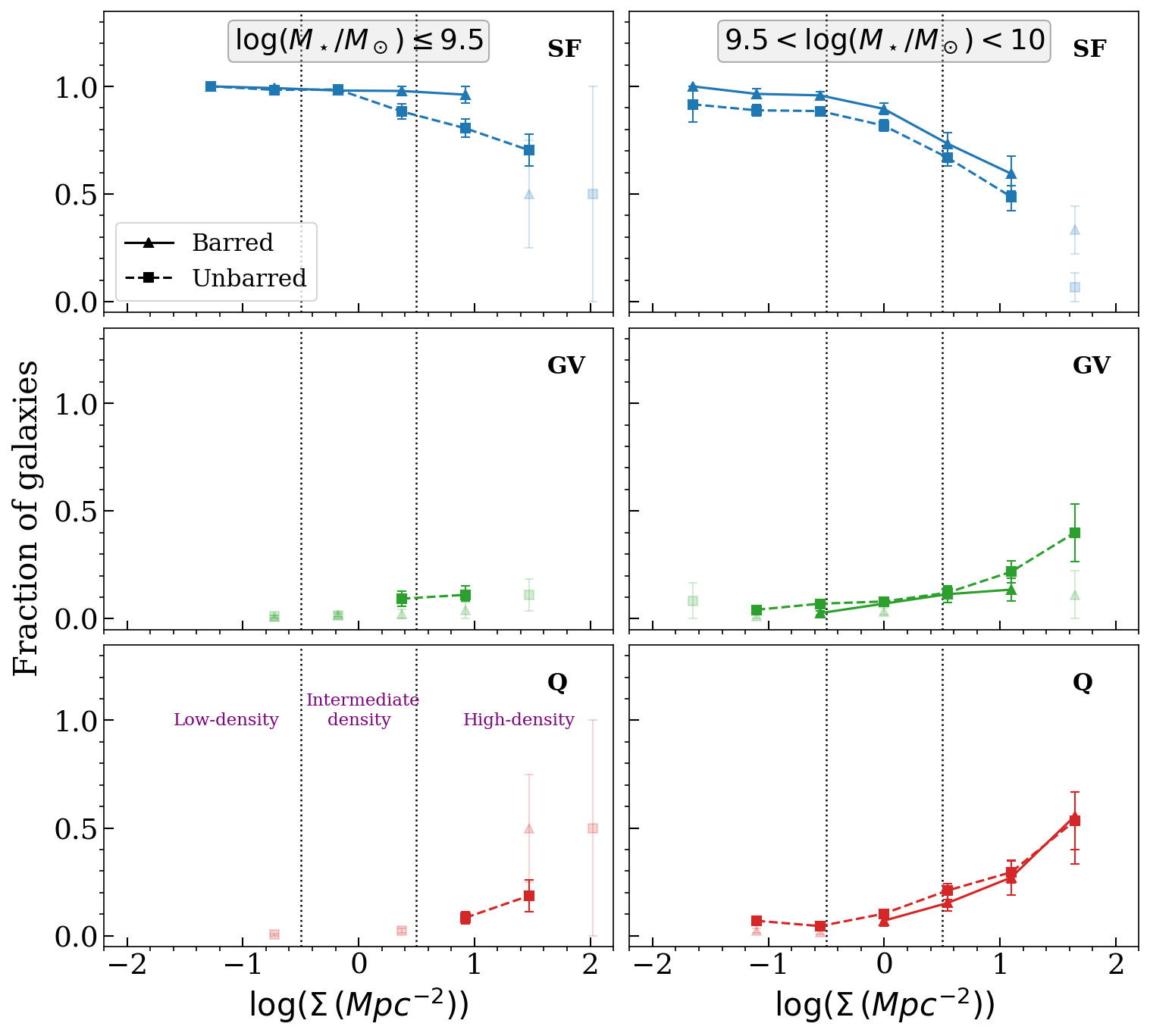}
    \end{subfigure}\hspace{-1.7mm}
    \begin{subfigure}[t]{0.59\textwidth}
        \centering
        \includegraphics[width=\linewidth]{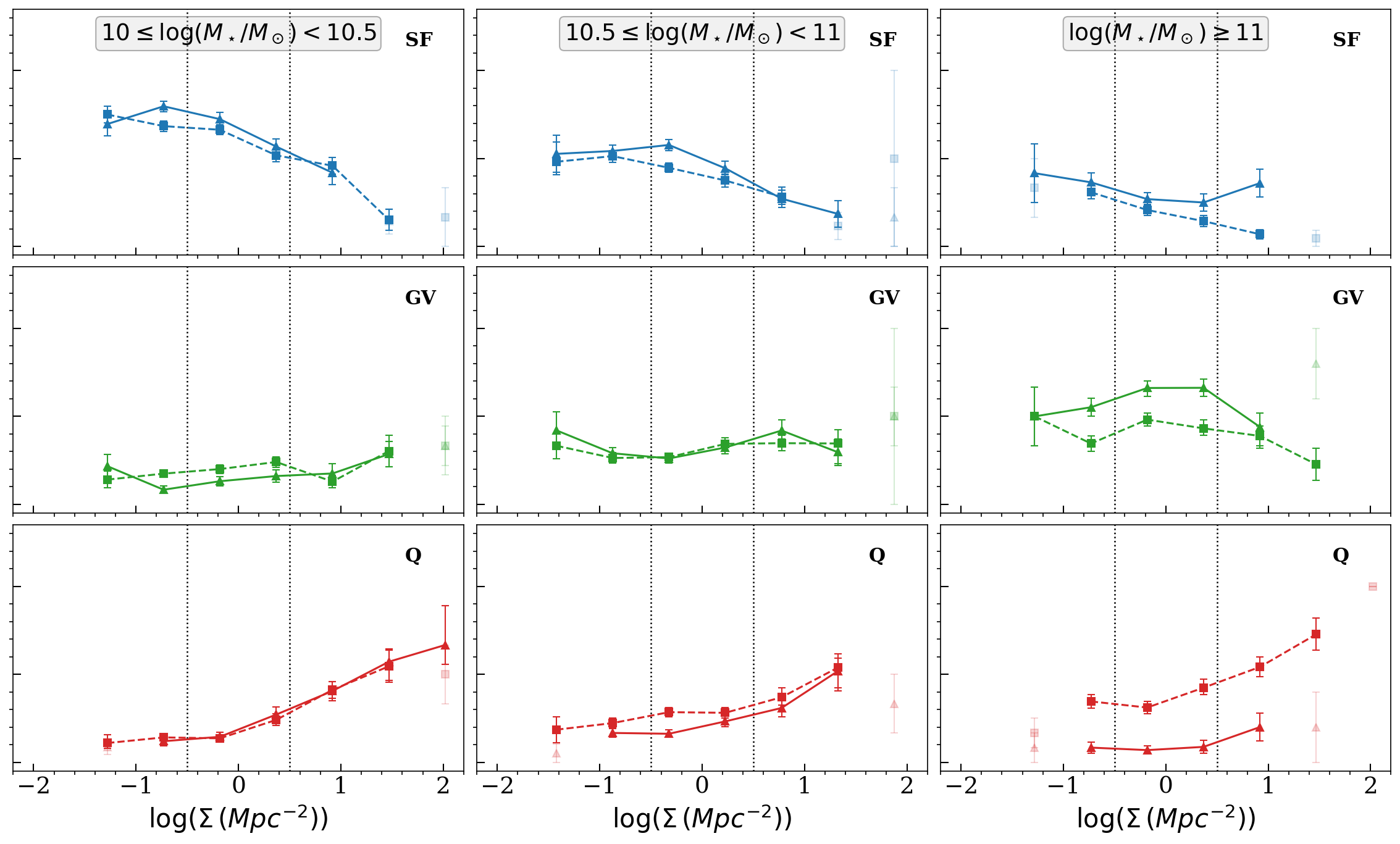}
    \end{subfigure}
    \caption{Distributions of the star-forming (SF), green valley (GV), and quenched (Q) galaxies as a function of local environmental density for barred (triangle) and unbarred (square) galaxies for different stellar mass bins. The rows correspond to SF (blue), GV (green), and Q (red). The vertical dotted lines divide the low-, intermediate-, and high-density environments. Bin size = 0.55 Mpc$^{-2}$. Faded points indicate bins with fewer than five galaxies. Error bars are estimated using bootstrap resampling.}
    \label{fig:bar_sf}
\end{figure*}

\clearpage
\bibliography{sample701}
\end{document}